\documentclass[12pt,preprint]{aastex}

\shorttitle{Effects of Neutral Particles on Modified Shocks}

\shortauthors{Ohira and Takahara}

\begin{document}

\title{Effects of Neutral Particles on Modified Shocks at Supernova Remnants}

\author{Yutaka Ohira\altaffilmark{1} and Fumio Takahara\altaffilmark{2}}

\begin{abstract}
H$\alpha$ emission from supernova remnants (SNRs) implies the existence of neutral hydrogens in the ambient medium. 
In the precursor of an SNR shock modified by cosmic rays (CRs), upstream plasmas are pushed by the CR pressure, but neutral particles are not, so that the relative velocity appears and some neutral particles become pickup ions by the charge exchange process in the precursor.
We investigate how the pickup ions generated in the precursor affect the shock structure and the particle acceleration.
If the CR pressure is larger than $20\%$ of the shock ram pressure, the compression of the subshock becomes smaller than that without pickup ions because of the pressure of the pickup ions.
Moreover, even if the shock is modified by CRs, the total compression ratio can be smaller than 4.
In addition, the pickup ions play an important role for the injection into the particle acceleration. 
If the shock is a quasi-perpendicular shock and if the the multiply reflected ion acceleration occurs, the CR spectrum can be harder than that of the test particle diffusive shock acceleration below GeV.    
\end{abstract}


\altaffiltext{1}{Theory Center, Institute of Particle and Nuclear Studies, KEK (High Energy Accelerator Research Organization), 1-1 Oho, Tsukuba 305-0801, Japan; ohira@post.kek.jp}
\altaffiltext{2}{Department of Earth and Space Science, Graduate School of Science, Osaka University, 1-1 Machikaneyama-cho, Toyonaka, Osaka 560-0043, Japan}
\section{Introduction}

Supernova remnants (SNRs) are observed in various bands from radio to gamma-ray. 
X-ray observations provide the evidence that electrons are accelerated to highly relativistic energies in SNR shocks \citep{koy95}. 
Moreover, it is believed that ions are also accelerated to the "knee" ($\sim10^{15}$ eV) of the Galactic cosmic ray (CR) spectrum providing most of the Galactic CRs. 
The most popular acceleration mechanism is diffusive shock acceleration (DSA) \citep{axford77,krymsky77,bell78,blandford78}.
To make the Galactic CRs, on average a few tens of percent of the explosion energy of a supernova has to be converted to the CR energy {so that} accelerated CRs change the shock structure to make the precursor in the upstream \citep{drury81}. 
On the other hand, the existence of neutral particles has been identified in many young SNRs from the detection of H$\alpha$ emissions. 
A recent review of H$\alpha$ emission from SNRs can be found in \citet{heng10}.
The neutral fraction of the interstellar matter around SNRs is often found to be order of unity. 
For example, \citet{gha00, gha02} have shown that the neutral fraction is about 0.8 and 0.1 in Tycho and SN1006 environments, respectively. 
Since SNR shocks are collisionless, neutral particles can penetrate the shock front freely without deceleration. 
As a result, a significant relative velocity exists between the plasma and neutral particles in the shock downstream region when the shock is unmodified by accelerated CRs. 
Through the charge exchange process in the downstream region, hot protons are converted to hot neutral hydrogens which emit broad H$\alpha$ emission \citep{chevalier78,chevalier80}.
Conversely, the cold neutrals are converted to cold protons which are not equilibrated with the downstream thermal protons due to the Coulomb collision because the relaxation time scale $t_{\rm pp}$
\begin{equation}
t_{\rm pp} \sim 3\times10^5~{\rm yr}\left(\frac{n_{\rm p}}{1~{\rm cm^{-3}}}\right)^{-1}\left(\frac{v_{\rm rel}}{3000~{\rm km~s^{-1}}}\right)^{3}~~,
\label{eq:pp}
\end{equation}
is much larger than the age of young SNR ($\sim 10^3~{\rm yr}$), where $n_{\rm p}$ and $v_{\rm rel}$ are the proton number density and the relative velocity between the cold protons and the thermal protons, respectively.
\citet{ohira09b} have shown that the charge-exchanged ions amplify the magnetic field and change the shock structure in the downstream.

In this letter, we propose that charge-exchanged ions are also important in the precursor of the shock modified by the CR because the relative velocity between hydrogens and protons exists in the precursor. 
Although the neutral hydrogen damps the plasma wave indispensable for DSA in the upstream region, the wave amplitude is still kept to be strong enough through the excitation by CRs near the shock front \citep{drury96, reville07}.
Hence, CRs can be accelerated in the partially ionized gas \citep{helder09,lee10}.
In fact, \citet{lee10} have observed H$\alpha$ from the precursor in the SNR Tycho.
The charge-exchanged ions excite plasma waves \citep{wu72,lee87,ohira09b} and are scattered by the waves \citep{williams94}. 
As the result, the plasma waves amplify the upstream magnetic field and the charge-exchanged ions become isotropic pickup ions. 
To estimate the efficiency of the charge exchange in the precursor, we compare the precursor length $\ell_{\rm pre}$ with the mean free path of the charge exchange $\ell_{\rm C. E.}=(n_{\rm p}\sigma_{\rm C. E.})^{-1}$, where $\sigma_{\rm C. E.}$ is the cross section of charge exchange which is almost constant $\sigma_{\rm C. E.}\sim10^{-15}{\rm cm}^2$ for $v_{\rm rel}<3000~{\rm km~s}^{-1}$ and a steeply decreasing function of $v_{\rm rel}$ for $v_{\rm rel}>3000~{\rm km~s}^{-1}$ \citep{heng07}. 
We here estimate the precursor length as the diffusion length of CRs $\ell_{\rm d}=D(p)/u_0$, where $p$, $D(p)$ and $u_0$ are the CR momentum, the diffusion coefficient of CR and the shock velocity, respectively.
Assuming the Bohm diffusion for CRs $D(p)=c^2p/3eB$ and $\sigma_{\rm C. E.}=10^{-15}~{\rm cm}^{2}$, we obtain
\begin{equation}
\frac{\ell_{\rm d}(p)}{\ell_{\rm C. E.}} = 3 \times 10^{-2} \left(\frac{p}{mc}\right) \left( \frac{n_{\rm p}}{1~{\rm cm}^{-3}} \right) \left( \frac{u_0}{3000~{\rm km~s}^{-1}} \right)^{-1} \left( \frac{B}{3~\mu{\rm G}} \right)^{-1}~~, 
\label{eq:efficiency}
\end{equation}
where $B$ is the magnetic field strength.
Thus, we expect that more than a few percent of hydrogens are converted to the pickup ions in the precursor.
Because on average $10^{-4-5}$ fraction of the upstream protons are needed to be injected into DSA to supply the Galactic CRs \citep{volk02}, the pickup ions are clearly important for the injection into CRs and for the modification of the shock structure.

We first provide a simple model of the modified shock structure (Section 2). We then solve the distribution function of the pickup ions (Section 3) and discuss impacts on the jump condition of subshock (Section 4) and the particle acceleration (Section 5).

\section{A simplified model of the modified shock structure in the precursor region}
\label{sec:2}
To calculate the distribution of pickup ions, we need to know the relative velocity between protons and hydrogens in the precursor.
In this section, we make a simplified model of the precursor structure of modified shock.
We assume that the pickup ions do not change the precursor structure, that is, the number of pickup ions is smaller than that of thermal protons.
The shock structure modified by the CR pressure has been calculated by various methods \citep{drury81,ellison90,berezhko97,blasi02,kang02}. 
In order to retain an essential feature of the structure of modified shocks, we determine the approximate structure of the precursor.
Mass and momentum conservation laws at the shock rest frame are 
\begin{eqnarray}
\rho_0 u_0 &=& \rho(x) u(x) \nonumber \\
\rho_0 u_0^2 + P_{\rm g,0} &=& \rho(x) u(x)^2 + P_{\rm g}(x) + P_{\rm CR}(x)~~,
\end{eqnarray}
where $\rho$, $u$, $P_{\rm g}$ and $P_{\rm CR}$ are the plasma density, flow velocity, the plasma pressure and the CR pressure, respectively.
Here, subscript $0$ means quantities far upstream and no subscript means those in the precursor, and $x$ is the coordinate of the direction along the shock normal ($x=0$ and $x=-\infty$ are the position of the subshock and far upstream, respectively.).
Neglecting the gas pressure, the velocity in the precursor is
\begin{equation}
\frac{u(x)}{u_0} = 1-\frac{P_{\rm CR}(x)}{\rho_0 u_0^2}~~,
\label{eq:u}
\end{equation}
where $P_{\rm CR}$ is given as
\begin{equation}
P_{\rm CR}(x) = \int_{mc}^{p_{\rm max}} \frac{1}{3}pcf_{\rm CR}(p,x)4\pi p^2dp~~,
\label{eq:pcr}
\end{equation}
where $f_{\rm CR}(p,x)$ is the distribution function of CR, and we assume that the pressure of non-relativistic CR is negligible. 
To make the expression simple, we use $f_{\rm CR}(p,x)$ as the approximate solution of test particle DSA \citep{blasi02},
\begin{equation}
f_{\rm CR}(p,x) \propto p^{-s}\exp{\left(\frac{x}{\ell_{\rm d}(p)}\right)}\sim p^{-s}\Theta(\ell_{\rm d}(p)+x)~~,
\label{eq:f}
\end{equation}
where $\Theta$ is the Heaviside step function. 
From Equations~(\ref{eq:u}), (\ref{eq:pcr}) and (\ref{eq:f}), we obtain the velocity structure of the plasma in the precursor 
\begin{eqnarray}
\frac{u(x)}{u_0}=1-\xi \times
\left \{ \begin{array}{ll}
\frac{\gamma_{\max}^{4-s}-(|x|/\ell_{\rm d,mc})^{4-s}}{\gamma_{\max}^{4-s}-1}&(\ell_{\rm d,\max}\ge |x| > \ell_{\rm d,mc})\\
1&(|x|\le  \ell_{\rm d,mc})
\label{eq:ux}
\end{array} \right. ~~,
\end{eqnarray}
where $\ell_{\rm d,mc}=\ell_{\rm d}(mc)$, $\xi=P_{\rm CR}(0)/\rho_0 u_0^2$, and $\gamma_{\max}$ is the maximum Lorentz factor of CRs.
We presume that $\xi$ is order of $0.1$ to be consistent with the general scheme of CR modified shock and the Galactic CR production.
Therefore, the velocity deceleration in the precursor region is around $u_0 - u(0)=\xi u_0 \sim 300~{\rm km~s}^{-1}$.
For $s<4$, the precursor scale $\ell_{\rm pre}$ is $\ell_{\rm d,\max} =\ell_{\rm d}(p_{\max})$ and for $s>4$,  $\ell_{\rm pre} \sim \ell_{\rm d,mc}$. 

\section{Distribution function of pickup ions}
\label{sec:3}
In this section, we calculate the distribution function of the pickup ions at the subshock, $f(v,x=0)$.
Charge-exchanged ions excite plasma waves and make a bispherical distribution by the pitch angle scattering with the excited waves. 
Here, we approximate the bispherical distribution as an isotropic distribution at plasma rest frame since the Alfv\'en Mach number ($M_{\rm A}$) of SNR shocks is high. 

From equation (\ref{eq:efficiency}), $\ell_{\rm pre}$ is usually smaller than $\ell_{\rm C. E.}$.
Then, pickup ions are produced in the whole precursor region and the flow velocity of hydrogens is $u_0$ in the precursor.
Hence, the relative flow velocity between the hydrogens and protons $\Delta u$ is
\begin{equation}
\Delta u(x) = u_0 - u(x)~~.
\end{equation}
We here assume that the thermal velocity of protons and hydrogens is smaller than $\Delta u$.
The validity of this assumption is discussed later.
The transit time of pickup ions over the precursor region is $\tau=\ell_{\rm pre}/(1-\xi)u_0$ which is less than $\ell_{\rm C. E.}/(1-\xi)u_0$, so that
\begin{equation}
\tau < 0.1~{\rm yr} \left(\frac{n_{\rm p}}{1{\rm cm^{-3}}}\right)^{-1}\left(\frac{1-\xi}{0.9}\right)^{-1}\left(\frac{u_0}{3000~{\rm km~s^{-1}}}\right)^{-1} ~~.
\end{equation}
From equation (\ref{eq:pp}), the equilibration time of the Coulomb collision in the precursor is
\begin{equation}
t_{\rm pp} = 3\times 10^2~{\rm yr}\left(\frac{n}{1~{\rm cm^{-3}}}\right)^{-1} \left(\frac{\xi}{0.1}\right)^3 \left(\frac{u_0}{3000~{\rm km~s^{-1}}}\right)^3~~.
\end{equation}
The recombination time $t_{\rm rec}$ is
\begin{equation}
t_{\rm rec} = 8\times10^4~{\rm yr}\left(\frac{n}{1~{\rm cm^{-3}}}\right)^{-1}\left(\frac{T_{\rm e}}{10^4~{\rm K}}\right)^{0.7}~~, 
\end{equation}
where we use the electron temperature $T_{\rm e}$ because the electron thermal velocity is comparable to or larger than $\Delta u$.
$t_{\rm pp}$ and $t_{\rm rec}$ are much larger than the transit time, so that we neglect the  collision and the recombination processes. 
Moreover, we neglect the charge exchange process between pickup ions and hydrogens because of $\ell_{\rm pre}<\ell_{\rm C. E.}$.
Hence, we consider only the charge exchange process between hydrogens and thermal protons. 

Then, the steady-state Boltzmann equation for pickup ions in the shock rest frame is
\begin{equation}
u(x) \frac{\partial f}{\partial x} = n_{\rm H}(x)n_{\rm p}(x)\sigma_{\rm C. E.}\Delta u(x)\frac{\delta(v-\Delta u(x))}{4\pi v^2} ~~,
\label{eq:boltzmann}
\end{equation}
where $v$ is the velocity of pickup ions at plasma rest frame and $n_{\rm H}$ is the hydrogen number density.
The solution is
\begin{equation}
f(v,x=0)=\frac{1}{4\pi v^2} \frac{u_0}{\ell_{\rm C. E.}}\int_{-\infty}^0 dx n_{\rm H}(x)\frac{\Delta u(x)}{u(x)^2}\delta(v-\Delta u(x))~~,
\label{eq:intf}
\end{equation}
where $\ell_{\rm C.E.}=(n_{\rm p,0}\sigma_{\rm C.E.})^{-1}$ and $n_{\rm p,0}$ is the proton number density far upstream. 
Then the distribution function of the pickup ions at the shock front is 
\begin{equation}
f(v,x=0) = f_1(v)+f_2(v)~~,
\end{equation}
where $f_1$ and $f_2$ are the distribution functions of the pickup ions ionized in the $x<-\ell_{\rm d,mc}$ and $x>-\ell_{\rm d,mc}$, respectively, and they are
\begin{equation}
f_1(v) = \frac{n_{\rm H}}{4\pi v^2}\frac{\ell_{\rm d,mc}}{\ell_{\rm C. E.}}\frac{\xi}{(1-v/u_0)^2}
\frac{\gamma_{\max}^{4-s}-1}{\xi u_0(4-s)}\left(\frac{v}{ \xi u_0}\right) \left\{ \gamma_{\max}^{4-s}-(\gamma_{\max}^{4-s}-1)\left(\frac{v}{\xi u_0}\right)\right \}^{-\frac{3-s}{4-s}}
\label{eq:f1}\\
\end{equation}
\begin{equation}
f_2(v) = \frac{n_{\rm H}}{4\pi v^2}\frac{\ell_{\rm d,mc}}{\ell_{\rm C. E.}} \frac{\xi}{(1-\xi)^2} \delta(v-\xi u_0)~~,
\label{eq:f2}
\end{equation}
where we approximate $n_{\rm H}$ as a constant. 
For $\gamma_{\max}\gg1$, $f_1$ is approximately expressed by
\begin{eqnarray}
f_1(v) &\simeq& \frac{n_{\rm H}}{4\pi v^2}\frac{\ell_{\rm d,\max}}{\ell_{\rm C. E.}} \frac{\xi}{(1-v/u_0)^2} \frac{1}{\xi u_0|4-s|} \left(\frac{v}{ \xi u_0}\right)\times \nonumber \\
&&~~~~~~~~~~~~~~~~\left \{ \begin{array}{ll}
\left\{ 1-\left(\frac{v}{\xi u_0}\right)\right \}^{-\frac{3-s}{4-s}}&(s<4)\\
\gamma_{\max}^{s-4} \left\{ 1+\gamma_{\max}^{s-4}\left(\frac{v}{\xi u_0}\right)\right \}^{-\frac{3-s}{4-s}} &(s>4)\\
\end{array} \right. ~~.
\label{eq:sf1}
\end{eqnarray}
%

\section{Effect of pickup ions on the jump condition of the subshock }
\label{sec:4}
In this section we discuss the jump condition of the subshock in the presence of pickup ions. 
As is well known for the termination shock of the solar wind ($u_0\sim400{\rm km~s^{-1}}$), large number of pickup ions change the jump condition of shocks \citep{richardson08}. 
The high pressure of pickup ions makes the Mach number and the compression ratio small. 
In addition, there is no guarantee that the behavior of the pickup ions in the shock is the same as that of gas with adiabatic index $5/3$ \citep{fahr08,wu09}.
Interestingly, the peculiar behavior (adiabatic index $> 5/3$) of pickup ions may allow the possibility that the total compression ratio can be smaller than 4 and the spectrum of CR can be softer than predicted by the test particle DSA, even if the shock is modified by the CRs.
The pressure of pickup ions at the subshock front is 
\begin{equation}
P_{\rm PI}=\frac{4\pi m}{3}\int_{0}^{\xi u_0}dvv^2f(v)=P_{\rm PI,1}+P_{\rm PI,2}~~,
\label{eq:pip}
\end{equation}
where $P_{\rm PI,1}$ and $P_{\rm PI,2}$ are partial pressures due to $f_1$ and $f_2$, respectively.
Approximating the third term of Equation (\ref{eq:sf1}) $\xi/(1-v/u_0)^2$ by $\xi/(1-\xi)^2$ , we obtain
\begin{eqnarray}
\frac{P_{\rm PI,1}}{\rho u^2} \simeq \frac{1}{3}\frac{\rho_{\rm H}}{\rho_0} \left(\frac{\xi}{1-\xi}\right)^3 \times
\left \{ \begin{array}{ll}
\frac{\ell_{\rm d,\max}}{\ell_{\rm C. E.}}\left(1-\frac{3}{5-s}+\frac{3}{9-2s}-\frac{1}{13-3s}\right) &(s<4)\\
\frac{\ell_{\rm d,mc}}{\ell_{\rm C. E.}}\gamma_{\max}^{13-3s}\left(\frac{1}{13-3s} -\frac{3}{9-2s}+\frac{3}{5-s}-1\right)&(4<s<13/3)\\
\frac{\ell_{\rm d,mc}}{\ell_{\rm C. E.}}\frac{1}{3s-13} &(s>13/3)\\
\end{array} \right. ~~,
\label{eq:p1}
\end{eqnarray}
where $\rho_{\rm H}$ is the density of hydrogens. 
From Equations (\ref{eq:f2}) and (\ref{eq:pip}), we obtain
\begin{equation}
\frac{P_{\rm PI,2}}{\rho u^2}=\frac{1}{3}\frac{\rho_{\rm H}}{\rho_0}\frac{\ell_{\rm d,mc}}{\ell_{\rm C. E.}} \left(\frac{\xi}{1-\xi}\right)^3~~.
\label{eq:p2}
\end{equation}
If the modification of the shock is not strong enough ($\xi<0.2$), the pressure of pickup ions is negligible ($P_{\rm PI}/\rho u^2<0.01$) , so that the effect of pickup ions on the precursor structure is negligible and our treatment in section \ref{sec:2} is valid.
If $s<13/3$, $\rho_{\rm H}\sim \rho_0$, $\xi \sim 0.5$ and if most hydrogens become the pickup ions in the precursor, from equation (\ref{eq:p1}) and (\ref{eq:p2}), the pressure of pickup ions account for about a few tens of percent of the ram pressure at the subshock, that is, the Mach number at subshock is almost $1$.
Therefore, the compression ratio at the subshock is almost $1$, so that the total compression ratio can be smaller than 4.
Although our treatment in section \ref{sec:2} is not valid in this case, we qualitatively can expect the reduction of subshock compression and the strong magnetic field amplification.

To calculate the jump condition of the subshock rigorously, we have to study the behavior of the pickup ions when they pass over the subshock \citep{matsukiyo07,wu09}. 
Although it should be investigated by particle in cell simulations, possible effects are mentioned in the next section.

\section{Effect of pickup ions on particle accelerations}
\label{seq:5}

As already discussed, the total compression ratio may become smaller than 4. Therefore, even when the SNR is young, that is, the Mach number of the shock is large, the spectral index of CRs may be softer than 2.
 
At present, the injection process for CR accelerations at shocks is still an open problem.
It is regarded that some particles above the threshold velocity are injected into the acceleration process.
As discussed above, the heating of plasma at the subshock is weak when the pickup ions drain a large fraction of the shock kinetic energy. 
Therefore, the pickup ions have the velocity much larger than that of thermal plasma and are important for the injection into CRs. 
This is a similar process to the solar wind termination shock where the pickup ions have been considered as the origin of the anomalous CR. 

\begin{figure}
\plotone{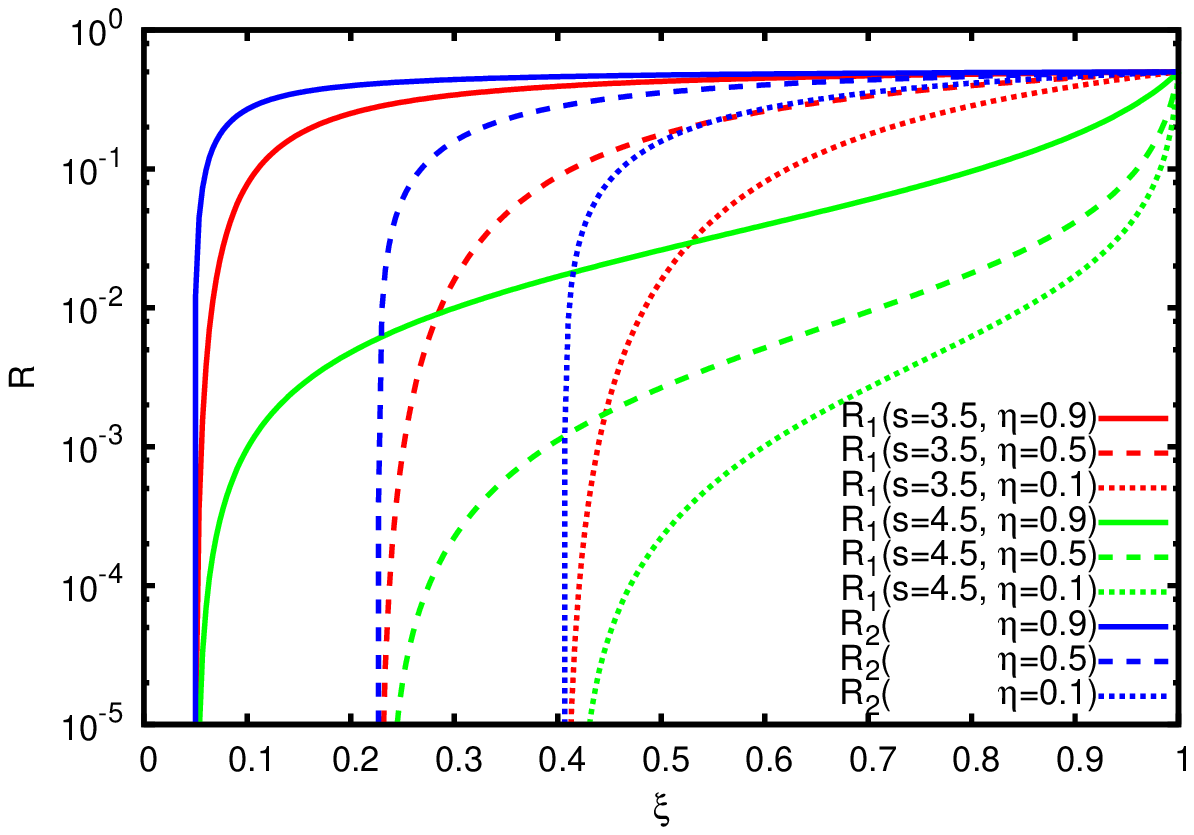} 
\caption{The reflection coefficient of the pickup ions. The red, green and blue lines show $R_1 (s=3.5)$, $R_1 (s=4.5,\gamma_{\max}=10^5)$ and $R_2$, respectively. The solid, dashed and dotted lines show $\eta=0.9$, $\eta=0.5$ and $\eta=0.1$, respectively.
\label{fig1}}
\end{figure}

The behavior of the pickup ions at the subshock of SNR has rarely been investigated. 
While the adiabatic acceleration is an important process, in this letter, we discuss the multiply reflected ion acceleration \citep{zank96,lee96} as another possible process.
When the shock is the collisionless quasi-perpendicular shock, there is an electrostatic potential to reflect some upstream thermal ions \citep{leroy83}.
The reflected thermal ions can penetrate the shock front after the reflection.
\citet{leroy83} estimated the amplitude of the electrostatic potential in the steady-state framework. However, for the high $M_{\rm A}$ shock, the shock is not stationary \citep{quest85} and the amplitude of time-averaged electrostatic potential is estimated by  
\begin{equation}
e\phi=\eta \frac{1}{2}mu^2~~,
\end{equation}
where $\eta$ is an order of unity parameter \citep{shimada05}. In the rest frame of the precursor plasma, particles satisfying $v_x<-A\xi u_0$ are reflected by the electrostatic potential, where $v_x$ is the velocity of the shock normal direction and $A=\xi^{-1}(1-\xi)(1-\eta^{1/2})$.
There are many pickup ions satisfying the reflection condition compared with thermal ions, and the pickup ions can be reflected by the electrostatic potential many times. 
The reflected ions drift to the direction of the motional electric field ${\bf u_0\times B}$.
As a result, the reflected ions are accelerated.
This acceleration mechanism is so-called shock surfing acceleration \citep{sagdeev66}.
The number of the pickup ions reflected by the potential $n_{\rm ref}$ is
\begin{equation}
n_{{\rm ref},i} = 2\pi \int_{A\xi u_0}^{\xi u_0} dv v^2f_i (v)\int_0^{\theta} d\theta' \sin \theta'~~,
\label{eq:ref}
\end{equation}
where $\theta$ is defined by $\cos \theta=A\xi u_0/v$, and $i = 1$ and $2$.
Since the total number density of the pickup ions $n_{{\rm PI},i}=2n_{{\rm ref},i}(A=0)$, from Equations (\ref{eq:f2}), (\ref{eq:sf1}), and (\ref{eq:ref}), the reflection coefficient of the pickup ions $R_i = n_{{\rm ref},i}/n_{{\rm PI},i}$ are 
\begin{eqnarray}
R_1 &=& \frac{1}{2} \times
\left \{ \begin{array}{ll}
(1-A)^{\frac{5-s}{4-s}}&(s<4)\\
\frac{1-A(5-s)-(s-4)(1+A\gamma_{\max}^{s-4})^{\frac{5-s}{4-s}}\gamma_{\max}^{5-s}}{1-(s-4)\gamma_{\max}^{5-s}}&(s>4)\\
\end{array} \right. \\
R_2 &=& \frac{1}{2}(1-A)~~.
\end{eqnarray}
Figure~\ref{fig1} shows the reflection coefficients where the red, green and blue lines show $R_1 (s=3.5)$, $R_1 (s=4.5, \gamma_{\max}=10^5)$ and $R_2$, respectively, the solid, dashed and dotted lines show $\eta=0.9$, $\eta=0.5$ and $\eta=0.1$, respectively. 
There exists the critical $\xi$ at which the reflection coefficient becomes 0. From the condition $A=1$, this critical $\xi_{\rm cr}$ is
\begin{equation}
\xi_{\rm cr}=\frac{1-\eta^{1/2}}{2-\eta^{1/2}}~~.
\end{equation}
When $\xi>\xi_{\rm cr}$, there are reflected pickup ions, so that it is possible that the multiply reflected ion acceleration occurs.
When the shock normal component of the Lorentz force becomes larger than the electric force of the electrostatic potential, the particle escapes to the downstream. 
So that the maximum velocity of accelerated particles is given by
\begin{equation} 
e\frac{v_{\max}}{c}B_{\perp} \sim e\frac{\phi}{L_{\rm ramp}}~~,
\label{eq:vmax}
\end{equation}
where $L_{\rm ramp}$ is the thickness of the electrostatic potential and $B_{\perp}$ is the magnetic field perpendicular to the shock normal direction.
The observation of the termination shock of the solar wind shows $L_{\rm ramp}\sim c/\omega_{\rm p,p}$ \citep{burlaga08}. 
From Equation~(\ref{eq:vmax}), the maximum velocity is
\begin{eqnarray} 
\frac{v_{\max}}{c} &\sim& \frac{\eta}{2}\left(\frac{u}{c}\right)^2\frac{\omega_{\rm p,p}}{\Omega_{\rm c,p}} \nonumber \\
&\sim&2\left(\frac{\eta}{1}\right) \left(\frac{u}{3000~{\rm km~s^{-1}}}\right)^2 \left(\frac{n_{\rm p}}{1~{\rm cm}^{-3}}\right)^{1/2} \left(\frac{B_{\perp}}{3~\mu \rm{G}}\right)^{-1}~~.
\end{eqnarray}
Therefore, the multiply reflected ion acceleration at SNR shocks can make relativistic CRs and is important for the injection mechanism. 
{This acceleration makes a harder spectrum than that of DSA \citep{zank96,lipatov99}, so that this may explain that the observed $\gamma$-ray spectrum from W51C is harder than expected for DSA \citep{abdo09}.}
However, since the high $M_{\rm A}$ ($M_{\rm A}>100$) collisionless shocks are not well understood, i.e., the shock structure may be different from the solar wind termination shock, it should be investigated whether the multiply reflected ion acceleration really occurs or not at SNR shocks.

\section{Discussion}

We discuss some important processes neglected in the above investigations.
If considerable hydrogens are converted to the pickup ions as discussed in section~\ref{sec:3}, they may change the structure of the precursor because of their large momentum flux. This issue will be addressed in future work.

In Section~\ref{sec:3}, we neglected any processes other than the charge exchange. 
%
%
If the upstream plasma is strongly heated by unstable waves in the precursor \citep{bell04,gha07,rakowski08,niemiec08,wagner09,ohira09a}, that is, if the temperature becomes $T > m (\xi u_0)^2$, neutral hydrogens are ionized by the electron scattering and the ionized particles can not be distinguished from thermal ions because thermal ions have the velocity comparable to that of the pickup ions.
\citet{rakowski09} showed that the collisional ionization occurs in the upstream of SNR.

In Section~\ref{sec:3}, we considered only the case $\ell_{\rm pre} < \ell_{\rm C. E.}$. 
If the magnetic field is not so strong and $s<4$, the precursor scale is the diffusion length of the maximum energy CR and $\ell_{\rm pre} > \ell_{\rm C. E.}$ (see equation (\ref{eq:efficiency}) and (\ref{eq:ux})).
For $\ell_{\rm pre} > \ell_{\rm C. E.}$, the multiple charge exchange is important  \citep{van08}. 
To take account of it and collisional ionization, we must  fully take into account the velocity dependence of the various atomic cross sections.
We will make full calculation in future work to investigate these issues.

\citet{raymond08} predicted that the pickup ions introduce a non-Gaussian broad H$\alpha$ line from downstream region.
We also predict it from upstream region.
So far, in calculations of H$\alpha$ from SNR, one has treated the pickup ions and thermal plasma as one fluid, and assumed that the pickup ions do not affect the plasma flow. 
As mentioned previously, the pickup ions should be treated as a separate component.

\section{Summary}

In this letter, we have investigated the effects of neutral particles on the shock modified by the CR at SNRs. 
More than a few percent of hydrogens are converted to pickup ions by the charge exchange in the precursor. 
In Section~\ref{sec:3}, assuming that the pickup ions become isotropic in the rest frame of the precursor plasma, we calculated the expected distribution of pickup ions at the precursor.
The pickup ions amplify the magnetic field and heat the upstream plasma. 
Moreover, the pickup ions have a large pressure in the precursor and change the jump condition of the subshock when the upstream plasma is strongly pushed by the CR pressure ($\xi= P_{\rm CR}(0)/\rho_0 u_0^2>0.2$). 
In addition, because of the large velocity of the pickup ions, they are preferentially accelerated by DSA. 
If the shock is the quasi-perpendicular shock and if the multiply reflected ion acceleration occurs, it is possible to make harder spectrum of CR than predicted by the test particle DSA below GeV.
Hence, the pickup ions are important for the injection into CR.

\acknowledgments
We thank T. Terasawa and M. Scholer for useful comments about pickup ions. 
We also thank the anonymous referee for valuable comments to improve the paper.
This work is partly supported by Scientific Research Grants (O.Y.: 21684014, and F.T.: 18542390 and 20540231) by the Ministry of Education, Culture, Sports, Science and Technology of Japan.

\clearpage

\end{document}